\documentclass[authoryear,preprint,12pt]{elsarticle}

\usepackage{epsfig}
\usepackage{epstopdf}
\usepackage{setspace}
\usepackage{microtype}
\usepackage{amssymb}
\usepackage{amscd}
\usepackage{amsmath}
\usepackage{amsfonts}
\usepackage[all,knot,poly]{xy}
\xyoption{arc}
\usepackage{latexsym}
\usepackage{psfrag} 
\usepackage{pstool} 

\usepackage[colorlinks=true,
pdfstartview=FitV, linkcolor=cyan,
citecolor=cyan, urlcolor=cyan]{hyperref}

\def\vector#1#2{\begin{vmatrix}
#1\vspace{4pt}\\#2
\end{vmatrix}}

\newcommand{\prova}{\par\noindent\textbf{Proof.} }
\newcommand{\fineprova}{\phantom{A}\hfill{$\blacksquare$}\goodbreak\medskip}

\newcommand{\LLc}{l}

\newcommand{\LLL}{\textrm{L}}

\newcommand{\kernel}{\phi}

\newcommand{\TANG}{T}

\newcommand{\conf}{{\BOmega}}

\newcommand{\di}[1]{(#1)}

\newcommand{\BOmega}{\boldsymbol{\Omega}}

\newcommand{\Bpi}{\boldsymbol{\pi}}

\newcommand{\Bss}{\boldsymbol{\sigma}}

\newcommand{\Bgamma}{\boldsymbol{\gamma}}

\newcommand{\Bn}{\mathbf{n}}
\newcommand{\Bo}{\mathbf{o}}
\newcommand{\Bp}{\mathbf{p}}

\newcommand{\Bu}{\mathbf{u}}
\newcommand{\Bv}{\mathbf{v}}

\newcommand{\BD}{\mathbf{D}}

\newcommand{\cM}{\mathcal{M}}

\newcommand{\VV}{V}

\newcommand{\LL}{L}

\newcommand{\TT}{T}

\newcommand{\MM}{M}
\newcommand{\NN}{N}

\newcommand{\yy}{y}

\newcommand{\scalar}[2]{{\langle}\kern.1em#1,#2\kern.1em{\rangle}}
\newcommand{\scalarC}[2]{{\langle}\kern.1em#1,#2\kern.1em{\rangle}_\conf}
\newcommand{\scalarT}[2]{{\langle}\kern.1em#1,#2\kern.1em{\rangle}_\TT}
\newcommand{\scalarTC}[2]{{\langle}\kern.1em#1,#2\kern.1em{\rangle}_{\TANG\conf}}

\newcommand{\integrale}[2]{\int_{#1}^{#2}}

\newcommand{\ee}{\varepsilon}

\newcommand{\sss}{s}
\newcommand{\mathscr}{\mathcal}

\newcommand{\sub}[1]{{}_{\lower2pt\hbox{$\scriptstyle#1$}}}

\newcommand{\punto}{\cdot}

\newcommand{\equaldef}{:=}

\newcommand{\set}[1]{\{\,#1\,\}}
\newcommand{\der}{d}

\newcommand{\norma}[1]{\| #1 \|}

\newcommand{\Eringen}{\persone{Eringen}}

\newtheorem{proposition}{Proposition}

\numberwithin{theorem}{section}
\numberwithin{proposition}{section}
\numberwithin{lemma}{section}
\numberwithin{remark}{section}
\numberwithin{definition}{section}

\def\Eringen{\textsc{Eringen}}

\def\c{\textit{c}}
\def\s{\textit{s}}
\def\t{\textbf{t}}
\def\tort{\t_{\bot}}
\def\k{\textbf{k}}

\def\R{\textbf{R}}

\input{tcilatex}

\usepackage[demo]{adjustbox}
\usepackage{xcolor}
\usepackage{atbegshi}
\AtBeginDocument{\AtBeginShipoutNext{\AtBeginShipoutDiscard}}	
\PassOptionsToPackage{hyphens}{url}\usepackage{hyperref} 

\begin{document}
	\pagestyle{empty} 
	\begin{titlepage}
		\color[rgb]{.4,.4,1}
		\hspace{5mm}

		\bigskip
		
		\hspace{15mm}
		\begin{minipage}{10mm}
			\color[rgb]{.7,.7,1}
			\rule{1pt}{226mm}
		\end{minipage}
		\begin{minipage}{133mm}
			\vspace{10mm}        
			\color{black}
			\sffamily
			\LARGE\bfseries On nonlocal mechanics of curved elastic beams  \\[-0.3\baselineskip]   \\[-0.3\baselineskip] 
			
			\vspace{5mm}
			{\large {Preprint of the article published in \\[-0.4\baselineskip] International Journal of Engineering Science \\[-0.1\baselineskip] 144, November 2019, 103140 }} 
			
			\vspace{10mm}        
			{\large Raffaele Barretta,\\[-0.4\baselineskip] \textsc{Francesco Marotti de Sciarra}, \\[-0.4\baselineskip] Marzia Sara Vaccaro} 
			
			\large
			
			\vspace{40mm}
			\vspace{5mm}
			
			\small
			\url{https://doi.org/10.1016/j.ijengsci.2019.103140}
			
			\textcircled{c} 2019. This manuscript version is made available under the CC-BY-NC-ND 4.0 license \url{http://creativecommons.org/licenses/by-nc-nd/4.0/}
			\hspace{30mm} 
			\color[rgb]{.4,.4,1} 
		\end{minipage}
	\end{titlepage}


\begin{frontmatter}



\title{On nonlocal mechanics of curved elastic beams}



\author[label1]{Raffaele Barretta}

\author[label1]{Francesco Marotti de Sciarra}
\author[label1]{Marzia Sara Vaccaro}

\address[label1]{
Department of Structures for Engineering and Architecture,\\ 
University of Naples Federico II, 
via Claudio 21, 80125 - Naples, Italy\\
e-mails: rabarret@unina.it - marotti@unina.it - marziasara.vaccaro@unina.it}

\begin{abstract}
Curved beams are basic structural components of Nano-Electro-Mechanical-Sistems (NEMS)
whose design requires appropriate modelling of scale effects.
In the present paper, the size-dependent static behaviour of curved elastic nano-beams 
is investigated by stress-driven nonlocal continuum mechanics. 
Axial strain and flexural curvature fields are 
integral convolutions between equilibrated axial force and bending moment fields 
and an averaging kernel.
The nonlocal integral methodology formulated here is the generalization 
to curved structures of the treatment in [Int. J. Eng. Science 115 (2017) 14-27] confined to straight beams. 
The corresponding nonlocal differential problem, supplemented
with non-standard boundary conditions, is highlighted and shown to lead to mathematically well-posed problems of nano-engineering.
The theoretical predictions, exhibiting stiffening nonlocal behaviours, are therefore
appropriate to significantly model a wide range of small-scale devices of nanotechnological interest.
The nonlocal approach is exploited by analytically establishing size-dependent responses
of curved elastic nano-sensors and nano-actuators that are driven by the small-scale characteristic parameter. 
\end{abstract}


\end{frontmatter}

\section{Introduction}
\label{sec:Intro}

Analysis, modelling, design, optimization and realization of smaller and smaller structural devices 
is nowadays a topic of major interest in Engineering Science, due to the recent advancements in Nanoscience and Nanotechnology.

\break
A large number of contributions have been published on size-dependent structural behaviour of rods, wires, beams, membranes, plates and shells at micro- and nano-scales to conceive modern small-scale systems, such as:  
varactors \citep{SedighiVaractor2017},
energy harvesters \citep{Lekha2017},
piezoelectric devices for biomedicine applications \citep{Salim2018},
resonators \citep{Chorsi2018},
resonant accelerometers \citep{Ding2019},
biosensors for detection of malaria protozoan parasites \citep{Kurmendra2019} and 
of cancer \citep{Kurmendra2019}, 
arch actuators \citep{OuakadSedighi2019}.
In this context, curved beams are the main structural elements of Micro/Nano-Electro-Mechanical-Systems
(M/NEMS) resonators 
\citep{Dantas2018,Alfosail2019,Frangi2019,NikpourianJCOMB2019,OuakadASME2019,%
Sieberer2019,Wang2019} 
that should be designed by significantly taking small size effects into account.
Only a few papers have been published on the matter and, however, scale phenomena 
have been modelled by strain-driven nonlocal or couple stress methods, 
which, as discussed below, present some operational difficulties.
\begin{enumerate}
\item
\textbf{Eringen differential nonlocal model}.
Such a strategy, recently employed in 
\citep{GanapathiPhysE2017,AyaJCOMB2017,PolitFEAD2018,RezaieeMAMS2018,ArefiJCOMB2019}, 
is based on the differential relation associated with the strain-driven fully nonlocal integral convolution of the theory of elasticity, originally exploited by \citet{Eringen1983} to tackle small-scale problems, defined in unbounded domains, of screw dislocation and wave propagation.
Such a model has been shown to provide, in the special case of straight structures, 
mechanical paradoxes 
\citep{Peddieson2003,ChallamelWang2008,LiYaoChenLiIJESci2015,Fernandez2016}
and size-dependent responses which are of limited applicative interest.
Inapplicability of Eringen's  strain-driven fully nonlocal integral model of elasticity to 
nanostructures has been highlited by
\citet{RB,Romano2017} and, nowadays, acknowledged by the scientific community of Engineering Science.
As a partial remedy, Eringen's strain-driven fully nonlocal integral model has been replaced with
the strain-driven local/nonlocal mixture \citep{Eringen1972,Eringen1987} 
to capture size effects
\citep{Wang2016,FernandezZaera2017,ZhuIJEsci2017,Khaniki2019,ZhangZAMM2019},  even if it
exhibits singular structural behaviors for vanishing local fractions
\citep{RomanoIntModels2017}.
\break
\item
\textbf{Nonlocal strain gradient model.}
This strategy, based on the differential law associated with the strain-driven nonlocal strain gradient integral model, was originally conceived by \citet{LimReddyJMPS2015} for applications in wave propagation.
In curved structures, the non-standard constitutive boundary conditions associated with the nonlocal gradient integral convolutions have been ignored 
\citep{SheIntJEngSci2019,SheActaAstro2019,KaramiIntJEngSci2019,%
SobhyJCOMB2019,EbrahimiMAMS2018,EbrahimiCompStruct2017}
and non-pertinent higher-order boundary conditions of strain gradient theory have been enforced.
This issue has been recently addressed by 
\citet{BarrettaNSGT,BarrettaVariationalNSGT} for straight nanobeams, and advantageously exploited by \citet{ApuzzoIJESci2018,ApuzzoJCOMB2019}.
\item
\textbf{Modified couple stress model.}
Such a strategy, proposed by \citet{Yang2002}, has been widely applied to small-scale structures 
\citep[see e.g.][]{DehrouyehSemnaniStability2017a,DehrouyehSemnaniStability2017b,FarokhiMotion2017,%
Ghayesh2017a,Ghayesh2017b,Farokhi2018a,Farokhi2018b,Farokhi2018c,%
GhayeshNonlinear2018,Khorshidi2018,TaatiIntJEngSci2018,%
Ghayesh2019,KaramiCS2019}.
Nevertheless, deformation measures 
describing couple stress theories are kinematically redundant,
as evidenced by \citet{RomanoRedundantCMAT} in the general framework of micromorphic continua.
Redundancy means that the assumed set of kinematical conditions ensuring rigidity of motion can be substituted with a proper subset of conditions.
To see this, we observe that, by virtue Euler's kinematic lemma \citep[][Lemma 1]{RomanoRedundantCMAT},
 vanishing of the macro-strain \citep[][Eq.(27)]{GhayeshFarajpour2019}
implies constancy and skew-symmetry of the displacement gradient.
This in turn implies that micro-spin \citep[][Eq.(31)]{GhayeshFarajpour2019}
is constant and that the microcurvature \citep[][Eq.(30)]{GhayeshFarajpour2019} is vanishing.
Adoption of the microcurvature tensor as an independent measure of deformation is therefore redundant
because its vanishing is already implied by vanishing of the macro-strain. 
The concept of kinematic redundancy of deformation measures, adopted in the context of generalized continua, was first introduced by \citet{RomanoRedundantCMAT} and then acknowledged by the
scientific community
\citep[see e.g.][]{BarbagalloIJSS2017,NeffPRoyal2017,SourkiEPJP2017}.
\end{enumerate}

\break

\section{Motivation and outline}
\label{sec:Motivation}

The stress-driven nonlocal integral elasticity approach of straight beams by \citet{RB2017}, 
effectively exploited for static, vibration, thermal and buckling problems of nanotechnological interest in 
\citep{AnsariActaMechSin2018,ApuzzoJCOMB2017,BarrettaMRC2018,%
BarrettaCanJCOMB2018,BarrettaCanIntJEngSci2018,BarrettaMAMS2019,%
BarrettaTimoEJMSol2018,BarrettaPlates,BarrettaBuckl},
is generalized in the present paper to curved nanobeams. 

The plan is the following.
Kinematics and equilibrium of Bernoulli-Euler curved beams
are preliminarily recalled in Sect.\ref{BEcin}.
The stress-driven nonlocal integral formulation of elasticity for curved structures 
is developed in Sect.\ref{sec:StrDrMod}.
Assuming that the integral model is governed by the special bi-exponential averaging kernel
adopted by \Eringen\ (\citeyear{Eringen1983}), an equivalent differential problem, with
homogeneous constitutive boundary conditions, is provided in
Proposition \ref{prop: nonlocal}.
The stress-driven nonlocal differential problem for straight beams \citep[][Prop.7.1, p.22]{RB2017}
is recovered as a special case by vanishing the geometric curvature of the nano-structural axis line. 
Equations governing the stress-driven nonlocal elastic equilibrium of curved beams are formulated in Sect.\ref{sec:NEP}.
Size-dependent elastic behavior of statically determinate and indeterminate curved beams
is established and evidenced by illustrations in
Sect.\ref{cases}. 
Results, discussion and closing remarks are summarized in Sect.\ref{sec:Concl}.

\section{Slender curved beams: kinematics and equilibrium}
\label{BEcin}

\begin{figure}[h]
\centering	
\includegraphics{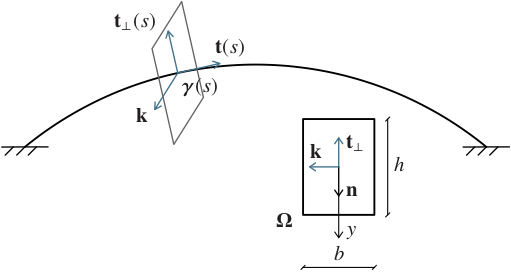}
\caption{Slender curved beam: a geometric sketch.}
\label{fig:beamsketch}
\end{figure}

Let us consider a slender curved beam of length $\,\LLL\,$ whose axis is detected
by a curve $\,\Bgamma:[0,\LLL]\mapsto\VV\,$, parametrized by the arc length $\,\sss\in[0,\LLL]\,$, 
in a plane $\,\Bpi\,$ endowed with the linear space of translations $\,\VV\,$.
The cross-section, assumed to be uniform along $\,\sss\,$, is modeled by a two-dimensional domain
$\,\Omega\,$. 
A geometric sketch of curved beam is depicted in fig.\ref{fig:beamsketch}.

\noindent
The field of tangent unit vectors to the beam axis is defined by 
\begin{equation}
\t \equaldef \Bgamma' 
\end{equation}
where the apex $\,(\bullet)'\,$ is the derivative with respect to $\,\sss\,$.
Denoting by $\,\R\,$ the linear operator which performs the rotation by $\,\pi/2\,$ counterclockwise 
in the plane $\,\Bpi\,$, we introduce the following field of transversal unit vectors
\begin{equation}
\tort\equaldef\R\t
\end{equation}
and the uniform unit vector field $\,\k\equaldef\t\times\tort\,$,
with $\,\times\,$ cross product.
According to Bernoulli-Euler's beam theory, cross-sections are clamped to the beam axis 
$\,\Bgamma\,$, and therefore the kinematic model is uniquely described by 
the translation velocity field $\,\Bv:[0,\LLL]\mapsto\VV\,$ of the beam axis.
Indeed, in this framework, the angular velocity field $\,\omega:[0,\LLL]\mapsto\Re\,$
of cross-sections is provided by the differential relation
\begin{equation}
\omega=\Bv'\punto\tort.
\label{fm:rotvel}
\end{equation}
The kinematically compatible deformation field $\,\BD_\Bv\,$
associated with the velocity field $\,\Bv:[0,\LLL]\mapsto\VV\,$ of the beam axis 
is given by 
\begin{equation}
\BD_\Bv
=
\vector{\ee_\Bv}{\chi_\Bv}
=
\vector{\Bv'\punto\t}{(\Bv'\punto\tort)'}
\label{fm:KinCompDef}
\end{equation}
in terms of axial strain $\ee_\Bv:[0,L]\mapsto\Re$ 
and flexural curvature $\chi_\Bv:[0,L]\mapsto\Re$ scalar fields.
The curved beam is subjected to an equilibrated force system
$\,f\,$ composed of a distributed vector loading $\,\Bp:[0,L]\mapsto\VV\,$ 
and of boundary concentrated forces
$\,{\mathbf{F}}_0\in\VV\,$ and $\,{\mathbf{F}}_L\in\VV\,$ and bending couples $\,\cM_0\in\Re\,$ and $\,\cM_L\in\Re\,$.
The equilibrated stress $\,\Bss\,$ in a Bernoulli-Euler curved beam
is the couple of axial force $\,N:[0,L]\mapsto\Re\,$
and bending moment $\,\MM:[0,L]\mapsto\Re\,$ fields
\begin{equation}
\Bss=
\vector{\NN}{\MM}
\end{equation}
fulfilling the virtual power condition
\begin{equation}
\setlength{\jot}{8pt}
\begin{aligned}
\scalar{f}{\delta\Bv}
=
\integrale{0}{L}
\Bss\di\sss
\punto
\BD_{\delta\Bv}\di\sss\,\der\sss
\,,
\label{fm:VWC}
\end{aligned}
\end{equation}
with
\begin{equation}
\setlength{\jot}{8pt}
\begin{aligned}
\scalar{f}{\delta\Bv}
\equaldef &
\integrale{0}{L}
\Bp\di\sss\punto\delta\Bv\di\sss\,\der\sss
+\cM_L\;\bigl(\delta\Bv'\di{\LL}\punto\tort\di\LL\bigr)
\\
&
+\cM_0\;\bigl(\delta\Bv'\di{0}\punto\tort\di{0}\bigr)
+{\mathbf{F}}_L\punto\delta\Bv\di\LL
+{\mathbf{F}}_0\punto\delta\Bv\di{0}\,,
\label{fm:Fsyst}
\end{aligned}
\end{equation}
virtual power of the force $\,f\,$,
for any virtual velocity field $\,\delta\Bv:[0,L]\mapsto\VV\,$
fulfilling homogeneous kinematic boundary conditions.
According to Eq.\eqref{fm:KinCompDef}, the deformation field $\,\BD_{\delta\Bv}\,$
kinematically compatible with the virtual velocity field $\,\delta\Bv\,$ is expressed by
\begin{equation}
\BD_{\delta\Bv}
=
\vector{\ee_{\delta\Bv}}{\chi_{\delta\Bv}}
=
\vector{{\delta\Bv}'\punto\t}{({\delta\Bv}'\punto\tort)'}
\,.
\label{fm:VirtKinCompDef}
\end{equation}
Integrating by parts Eq.\eqref{fm:VWC}, applying a standard localization procedure
and projecting along the axial $\,\t\,$ and transversal $\,\tort\,$ directions
we get the equivalent equilibrium differential problem
in $\,[0,\LLL]\,$
\begin{equation}
\setlength{\jot}{8pt}
\left\{
\begin{aligned}
\NN'+\TT\,(\tort'\punto\t)
&=-\Bp\punto\t\,,
\\
\TT'+\NN\,(\t'\punto\tort)
&=-\Bp\punto\tort\,,
\label{fm:DiffProbl}
\end{aligned}
\right.
\end{equation}
equipped with the boundary conditions
at the abscissae $\,\sss=0\,$ and $\,\sss=\LL\,$
\begin{equation}
\setlength{\jot}{8pt}
\left\{
\begin{aligned}
-\NN\di{0}\punto(\delta\Bv\punto\t)\di{0}
&=({\mathbf{F}}_0\punto\t\di{0})\punto(\delta\Bv\punto\t)\di{0}\,,
\\
\NN\di{L}\punto(\delta\Bv\punto\t)\di\LL
&=({\mathbf{F}}_L\punto\t\di\LL)\punto(\delta\Bv\punto\t)\di\LL\,,
\\
-\MM\di{0}\;\delta\omega\di{0}
&=\cM_0\;\delta\omega\di{0}\,,
\\
\MM\di{L}\;\delta\omega\di{\LL}
&=\cM_L\;\delta\omega\di{\LL}\,,
\\
-\TT\di{0}\punto(\delta\Bv\punto\tort)\di{0}
&=({\mathbf{F}}_0\punto\tort\di{0})\punto(\delta\Bv\punto\tort)\di{0}\,,
\\
\TT\di{L}\punto(\delta\Bv\punto\tort)\di\LL
&=({\mathbf{F}}_L\punto\tort\di\LL)\punto(\delta\Bv\punto\tort)\di\LL
\,,
\label{fm:StatBC}
\end{aligned}
\right.
\end{equation}
with $\,\TT\equaldef-\MM':[0,\LLL]\mapsto\Re\,$ shear force interaction field.

In many engineering problems the beam axis is generally regular, so that the  
vector $\,\t'\,$ is not vanishing and leads to the definition of 
normal unit vector $\,\Bn\equaldef\t'/c\,$, with
$\,c\equaldef\norma{\t'}\,$ geometric curvature of the beam axis. 
The relationship between $\,\tort\,$ and $\,\Bn\,$ is given by \citet{RomTomoUno}: 
$\,\tort=(\tort\punto\Bn)\,\Bn\,$.

\section{Stress-driven nonlocal integral elasticity}
\label{sec:StrDrMod}

Let us preliminarily recall the local elasticity linearized model of Bernoulli-Euler curved beam.
Local elastic axial strains $\,\ee\,$ and flexural curvatures $\,\chi\,$ are related to
axial forces $\,N\,$ and bending moments $\,M\,$ by
\cite{Winkler1858,Baldacci}
\begin{equation}
\setlength{\jot}{8pt}
\left\{
\begin{aligned}
\ee\di{s} &= \biggl[\frac{1}{EA}[N - \c\, M]\biggr]\di{s}   \,,
\\
\chi\di{s} &= \biggl[
\frac{M}{EJ_r} - \frac{\c}{EA}[N - \c\, M]\biggr]\di{s}\,,
\end{aligned}
\right.
\label{LocEq}
\end{equation}
with $\,E\,$ Euler-Young modulus, $\,A\,$ cross-sectional area 
and $\,J_r\,$
is the reduced moment of inertia 
\begin{equation}
J_r
\equaldef
\integrale{\Omega}{}\yy^2\frac{1}{1-c\,\yy}\,\der A\,,
\label{fm:RedMomIner}
\end{equation}
along the bending axis $\,\yy\,$ associated with unit normal vector $\,\Bn\,$.

Nonlocal effects are described by requiring that elastic axial strains and flexural curvatures
are integral convolutions between an averaging kernel $\,\kernel_\LLc:\Re\mapsto[0;+\infty[\,$, 
described by a nonlocal length parameter $\,\LLc\in]0;+\infty[\,$, and the local fields
formulated in Eq.\eqref{LocEq}.
The stress-driven nonlocal integral model takes thus the form
\begin{equation}
\setlength{\jot}{8pt}
\left\{
\begin{aligned}
\ee\di{s}&= \int_{0}^{L}{\kernel_{\LLc}(s-\xi) \: \biggl[\frac{1}{EA}[N - \c\,M]\biggr]\di{\xi}\:d\xi}    
\\
\chi\di{s} &= \int_{0}^{L}{\kernel_{\LLc}(s-\xi) \: \biggl[\frac{M}{EJ_r} - \frac{\c}{EA}[N - \c\,M]\biggr]\di{\xi}\:d\xi}
\end{aligned}
\right.
\label{StressConv}
\end{equation}
Equipping Eq.\eqref{StressConv} with the special bi-exponential kernel in \citep{Eringen1983}
\begin{equation}
\kernel_{\LLc}\di{s} = \frac{1}{2\LLc} \exp(-\frac{|s|}{\LLc})\,,
\label{kern}
\end{equation}
the stress-driven integral convolutions Eq.\eqref{StressConv} can be 
inverted to give the following constitutive equivalence result, extending thus to curved beams the 
formulation in \citep[][Prop.7.1, p.22]{RB2017} confined to straight beams. 
The proof is analogous. 
\begin{proposition}\label{prop: nonlocal}
The nonlocal integral formulation for elastic curved beams Eq.\eqref{StressConv},
equipped with the bi-exponential kernel Eq.\eqref{kern},
is equivalent to the differential problem
\begin{equation}
\setlength{\jot}{8pt}
\left\{
\begin{aligned}
&\ee\di{s}-\LLc^2\,(\partial_s^2\ee)\di{s}
=\biggl[\frac{1}{EA}[N - \c\, M]\biggr]\di{s}
\,,
\\
&\chi\di{s}-\LLc^2\,(\partial_s^2\chi)\di{s}
=\biggl[
\frac{M}{EJ_r} - \frac{\c}{EA}[N - \c\, M]\biggr]\di{s}\,,
\end{aligned}
\right.
\label{fm: Stressdifform}
\end{equation}
with the constitutive boundary conditions (CBC)
\begin{equation}
\setlength{\jot}{8pt}
\left\{
\begin{aligned}
&\partial_s \ee\di{0}=\frac{1}{\LLc}\,\ee\di{0}\,,
\\
&\partial_s \ee\di{L}=-\frac{1}{\LLc}\,\ee\di{L}\,,
\\
&\partial_s\chi\di{0}=\frac{1}{\LLc}\,\chi\di{0}\,,
\\
&\partial_s\chi\di{L}=-\frac{1}{\LLc}\,\chi\di{L}\,.
\end{aligned}
\right.
\label{fm: StressEqui}
\end{equation}
\end{proposition}

\section{Stress-driven nonlocal elastic equilibrium of curved beams}
\label{sec:NEP}

The elastostatic problem of a Bernoulli-Euler curved beam is formulated in terms
of displacement field $\,\Bu:[0,\LLL]\mapsto\VV\,$ of the structural axis. 

In the realm of the geometrically linearized theory, the kinematic relation Eq.\eqref{fm:KinCompDef}
holds true, so that the deformation field $\,\BD_\Bu\,$ of the curved nanobeam,
associated with the displacement field $\,\Bu:[0,\LLL]\mapsto\VV\,$, writes as 
\begin{equation}
\BD_\Bu
=
\vector{\ee_\Bu}{\chi_\Bu}
=
\vector{\Bu'\punto\t}{(\Bu'\punto\tort)'}
\label{fm:KinCompDefSpost}
\end{equation}
in terms of linearized axial strain $\ee_\Bu:[0,L]\mapsto\Re$ 
and linearized flexural curvature $\chi_\Bu:[0,L]\mapsto\Re$ scalar fields.

Resorting to Eq.\eqref{fm:rotvel}, the field of rotations 
$\,\varphi:[0,\LLL]\mapsto\Re\,$ of cross-sections
is obtained by integrating Eq.\eqref{fm:KinCompDefSpost}$_2\,$
\begin{equation}
\varphi\di\sss=(\Bu'\punto\tort)\di\sss
=\varphi\di{0}
+
\integrale{0}{\sss}
\chi_\Bu\di\xi
\,\der\xi\,.
\label{fm:rot}
\end{equation}
To analytically establish the expression of the vector displacement field, 
let us preliminarily consider the additive decomposition
\begin{equation}
\Bu'
=(\Bu'\punto\t)\,\t
+(\Bu'\punto\tort)\,\tort
=\ee_\Bu\,\t
+\varphi\,\tort\,,
\label{fm:decomp}
\end{equation}
where Eqs.\eqref{fm:KinCompDefSpost} and \eqref{fm:rot} are enforced.

The sought displacement field of the curved nanobeam is obtained by 
integrating Eq.\eqref{fm:decomp}
\begin{equation}
\Bu\di\sss
=\Bu\di{0}
+\integrale{0}{\sss}\Bu'\der\xi\,
=\Bu\di{0}
+
\integrale{0}{\sss}
[\ee_\Bu\di\xi\,\t\di\xi
+\varphi\di\xi\,\tort\di\xi]
\,\der\xi\,.
\label{fm:spost}
\end{equation}
The unknown integration constants $\,\varphi\di{0}\in\Re\,$ and $\,\Bu\di{0}\in\VV\,$
are univocally evaluated by prescribing the essential kinematic boundary conditions at hand.
In all the case-studies presented in the next section, 
the total deformations $\,\set{\ee_\Bu,\chi_\Bu}\,$ are assumed to be coincident with the
elastic deformation fields $\,\set{\ee,\chi}\,$ provided 
by the stress-driven nonlocal integral convolutions Eq.\eqref{StressConv}.

\section{Case-studies}
\label{cases}

The developed stress-driven methodology is illustrated by investigating 
the nonlocal behaviour of four static schemes of nanotechnological interest.
The nanobeam axis is assumed to be a circle arc of radius $\,r= 10\,$ [nm], so that 
length and curvature are given by $\,L = \pi  / (2\c)\,$ and $\,\c=1/r\,$, respectively.  
The cross-section $\,\Omega\,$ is a rectangle of base $\,b = 1\,$ [nm] and height $\,h = 2\,$ [nm].
The reduced moment of inertia is evaluated by Eq.\eqref{fm:RedMomIner}
\begin{equation}
J_r = \int_{-b/2}^{+b/2}{\int_{-h/2}^{+h/2}{\frac{r \: y^2}{r - y}\:dy}\:dx} 
\label{inerzia}
\end{equation}
The nanobeam is assumed to be made of silicon carbide, with Euler-Young modulus 
$\, E=427\,$ [GPa].

\noindent
In the illustrations below, the following dimensionless nonlocal parameter
\begin{equation}
\lambda = \frac{\LLc}{L}\,,
\label{equ13}
\end{equation}
with $\,\LLc\,$ characteristic length parameter and $\,L\,$ nanobeam length, is adopted.

\begin{figure}[htp]
\centering	
\includegraphics{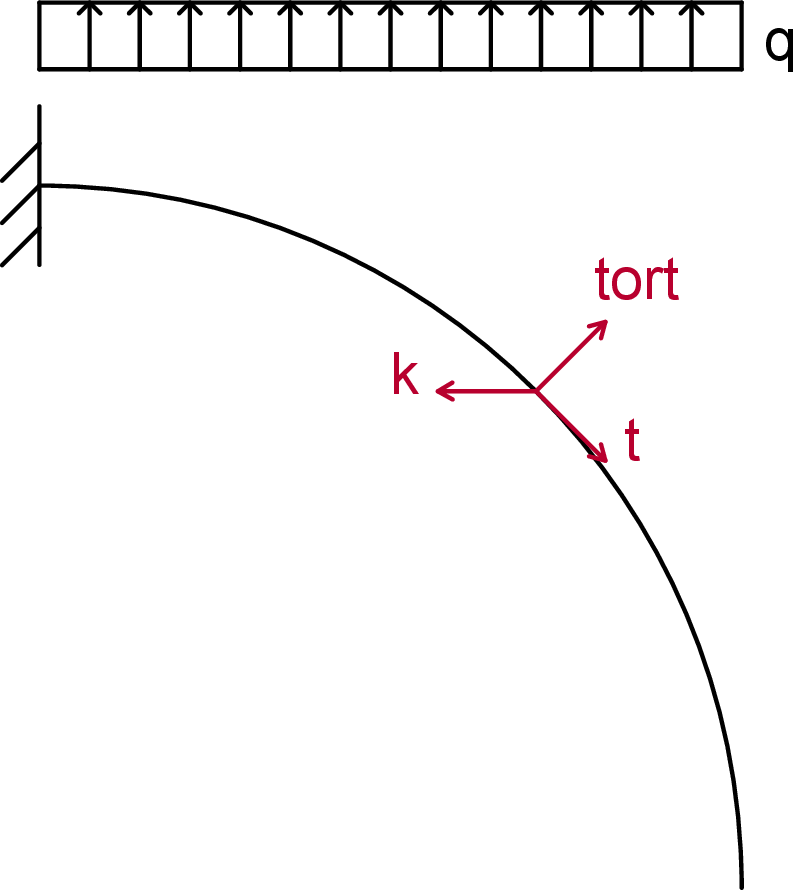}
\caption{Cantilever curved nanobeam.}
\label{fig0}
\end{figure}

\begin{figure}[htp]
\centering	
\includegraphics{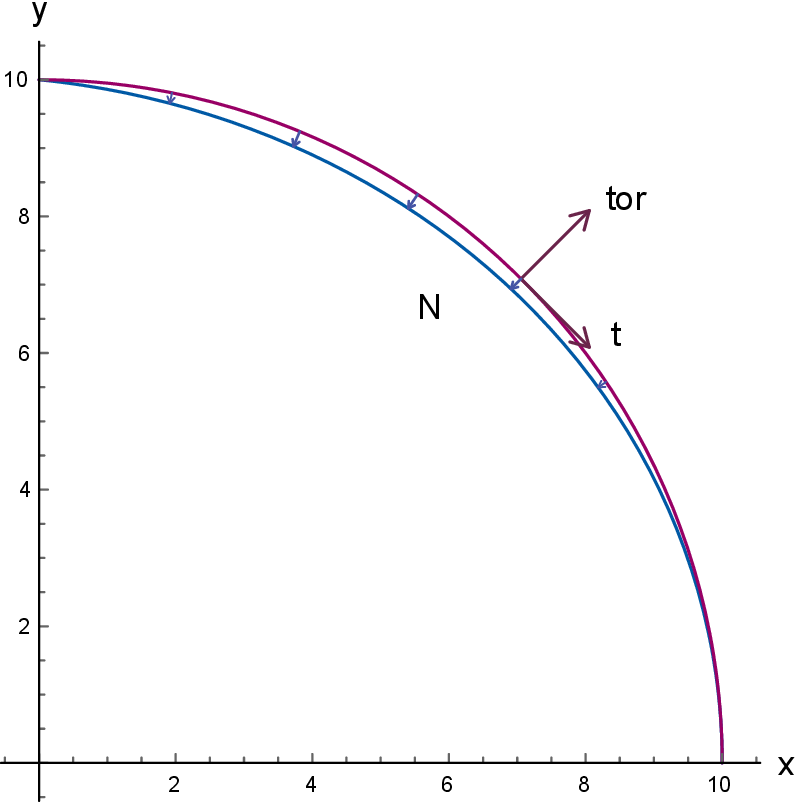}
\caption{Cantilever curved nanobeam: plot of the vector field $N\tort$.}
\label{fig2}
\end{figure}

\begin{figure}[htp]
\centering	
\includegraphics{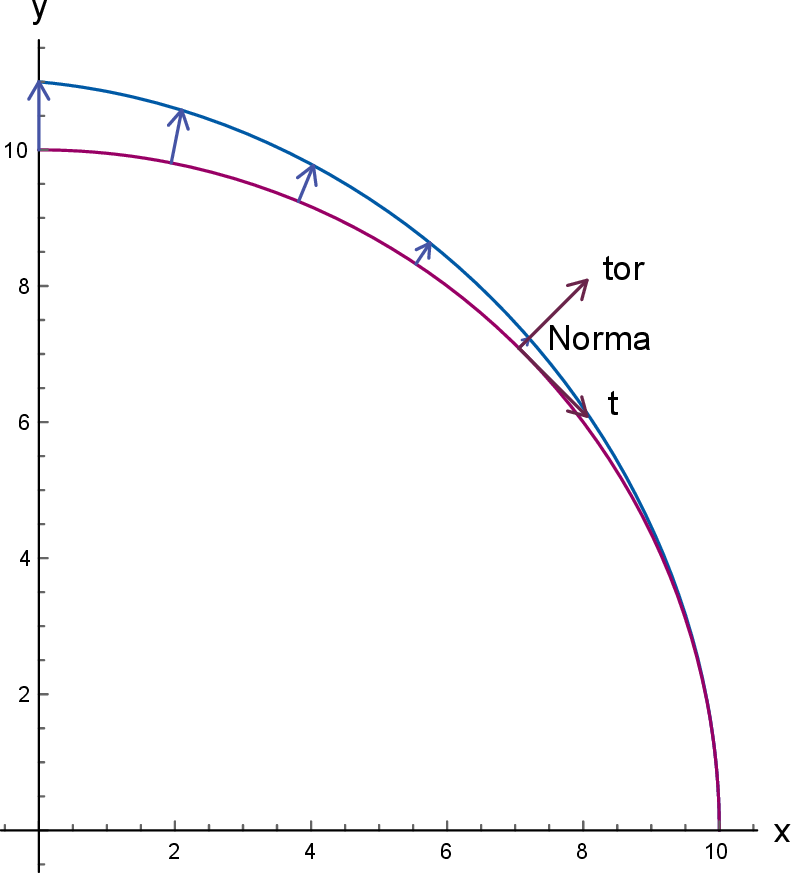}
\caption{Cantilever curved nanobeam: plot of the vector field $T\tort$.}
\label{fig3}
\end{figure}

\begin{figure}[htp]
\centering	
\includegraphics{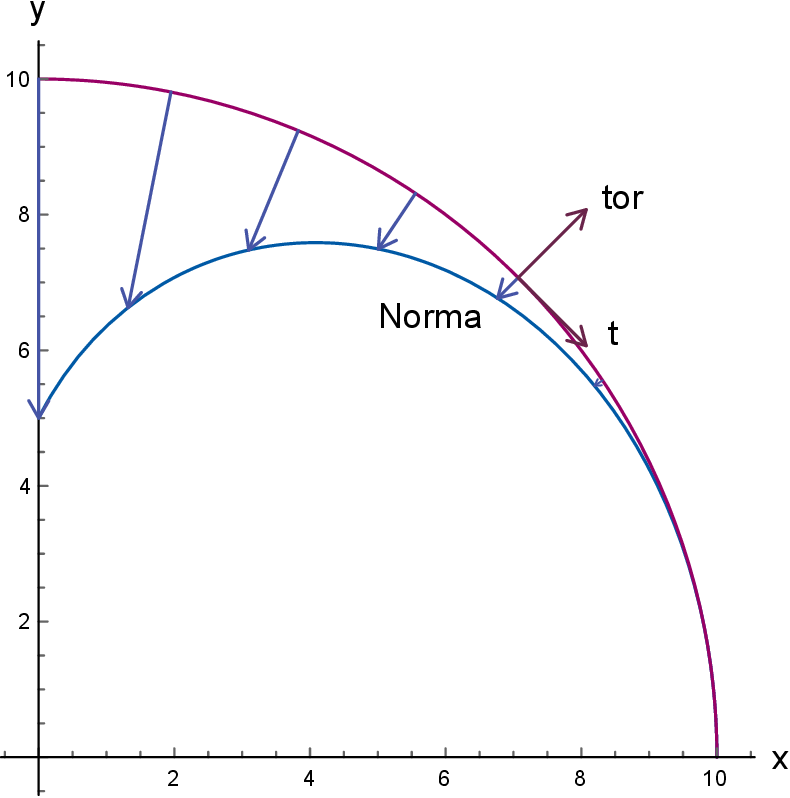}
\caption{Cantilever curved nanobeam: plot of the vector field $\R ($M$\k) = -M\tort$.}
\label{fig4}
\end{figure}

\begin{figure}[htp]
\centering	
\includegraphics{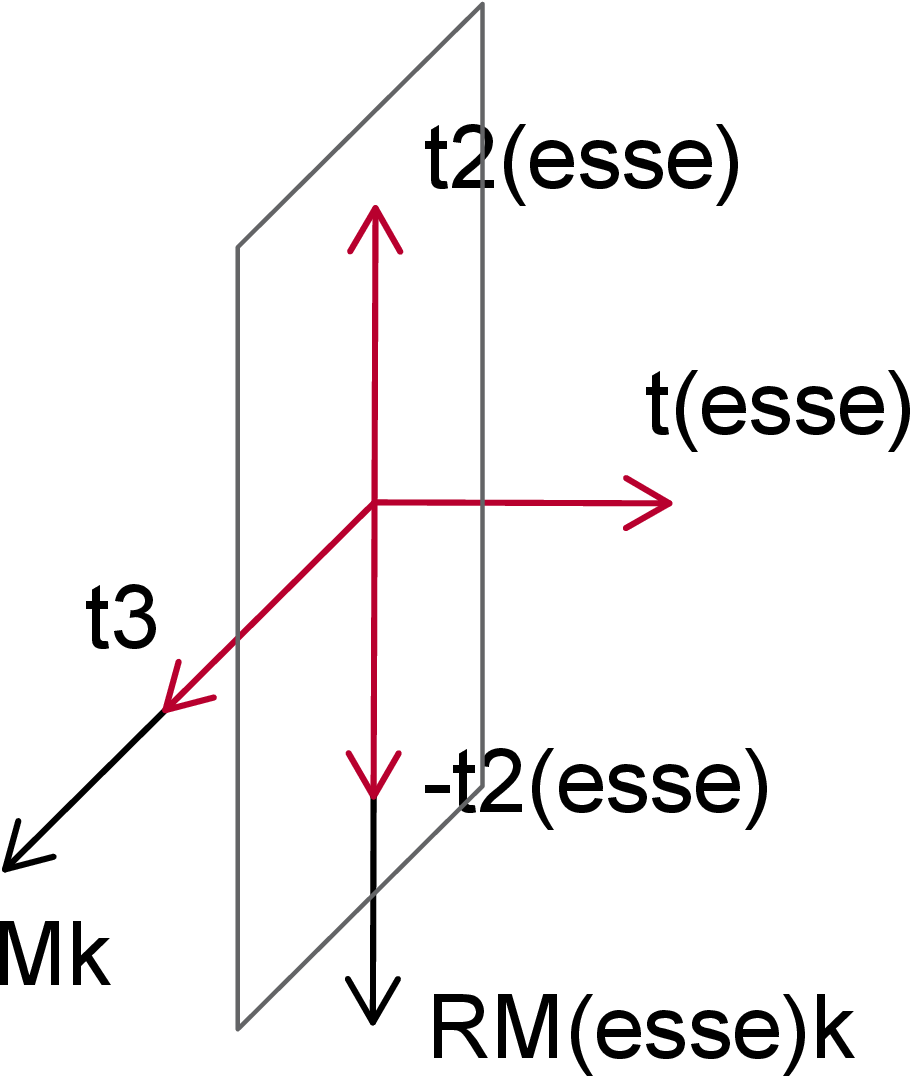}
\caption{Cross section at abscissa $\,\s\,$ with the local triad and the vectors $\,M\k\,$ and $\,\R M\k\,$.}
\label{cross_sec}
\end{figure}

\begin{figure}[htp]
\centering	
\includegraphics{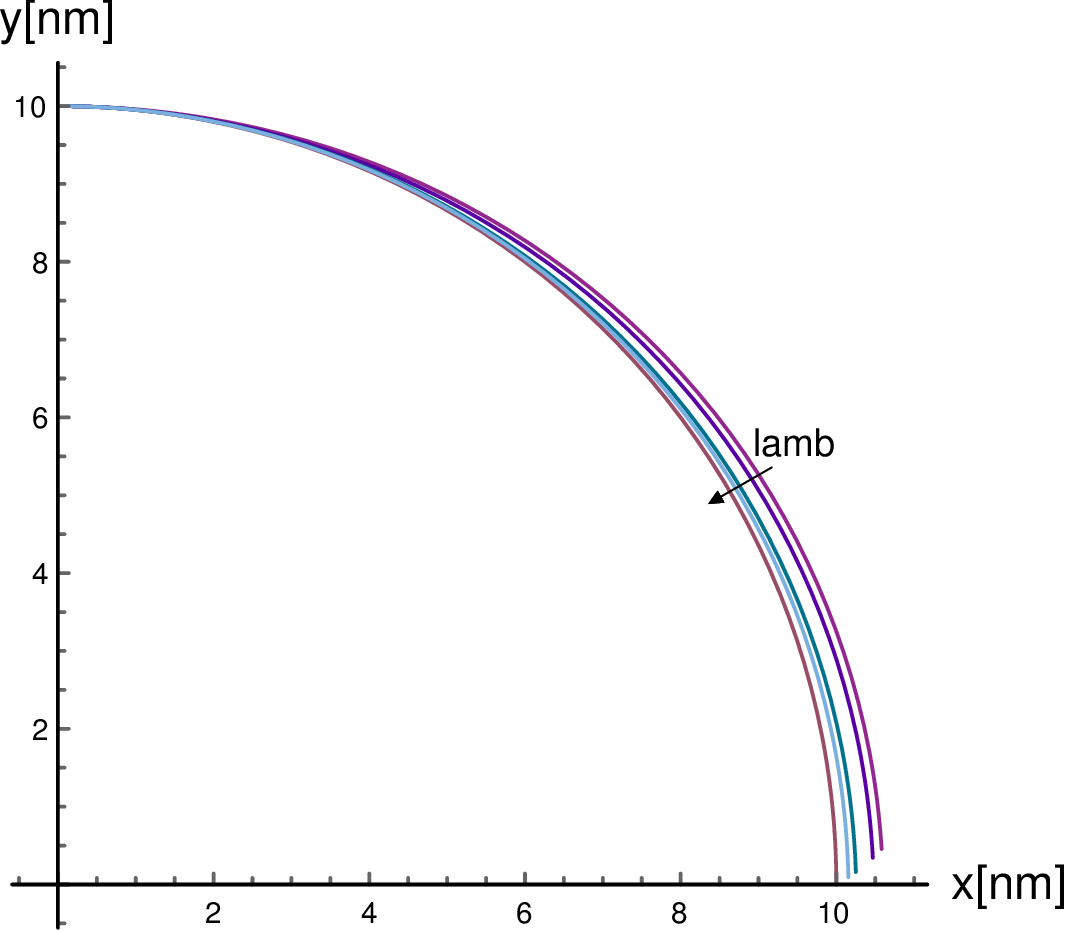}
\caption{Cantilever curved nanobeam: deflection of the beam axis for $\lambda\in \{0; 0.1; 0.5; 1\}$.}
\label{fig1}
\end{figure}

\begin{figure}[htp]
\centering	
\includegraphics{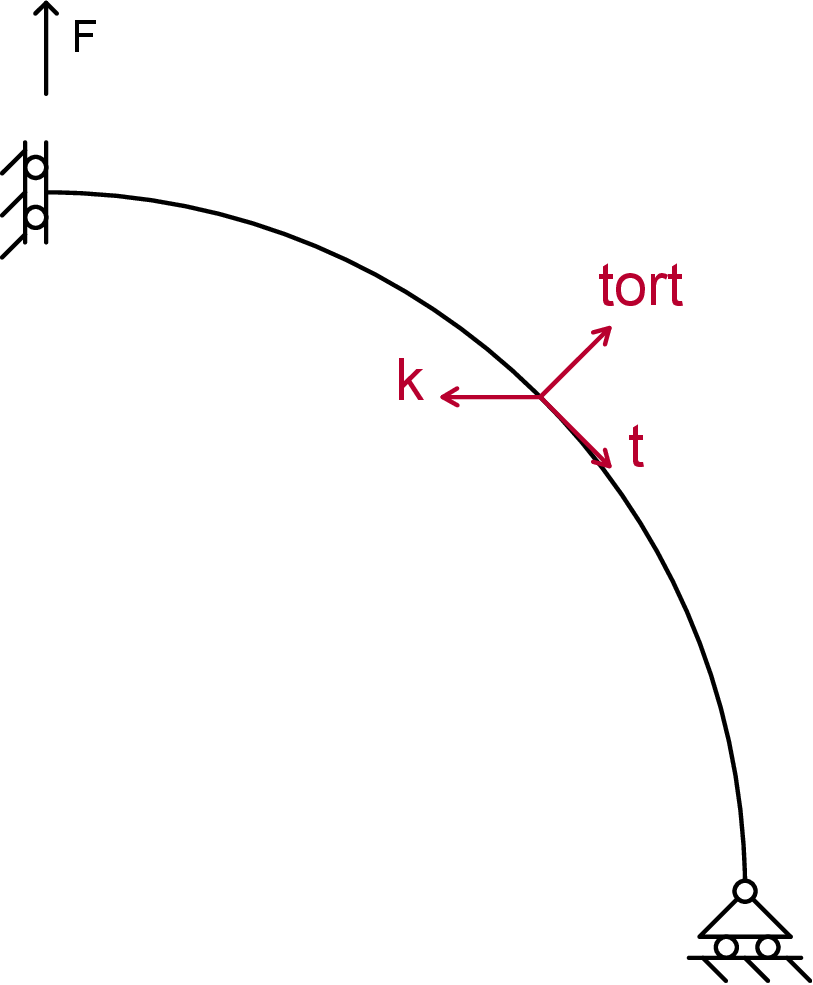}
\caption{Curved nanobeam with slider and roller supports.}
\label{fig0a}
\end{figure}

\begin{figure}[htp]
\centering	
\includegraphics{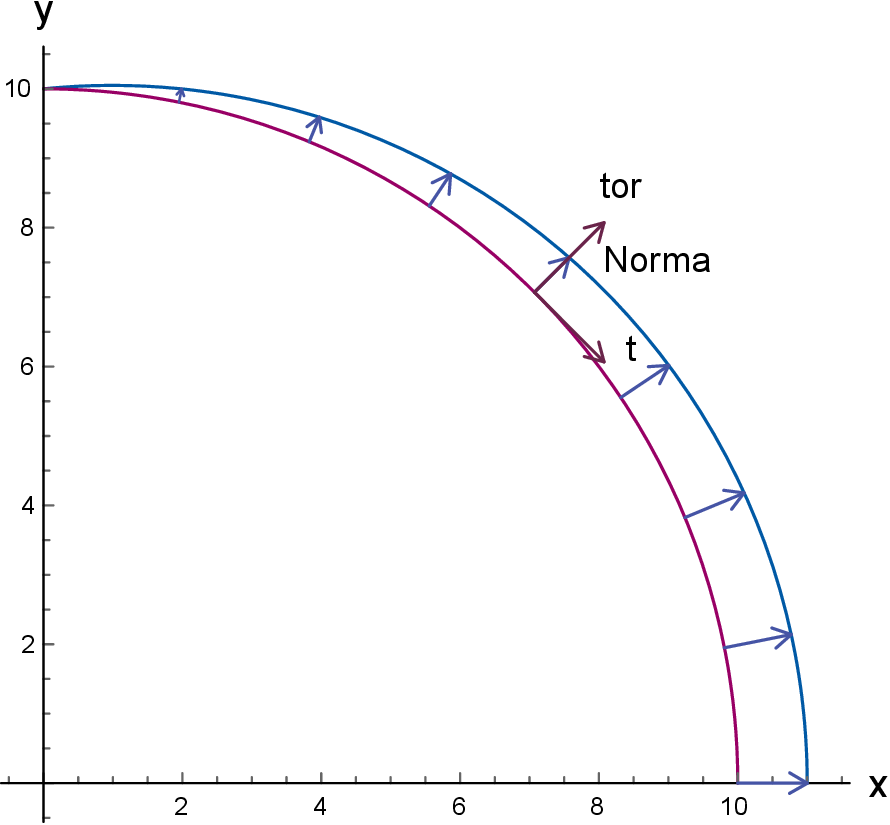}
\caption{Curved nanobeam with slider and roller supports: plot of the vector field $N\tort$.}
\label{fig5}
\end{figure}

\begin{figure}[htp]
\centering	
\includegraphics{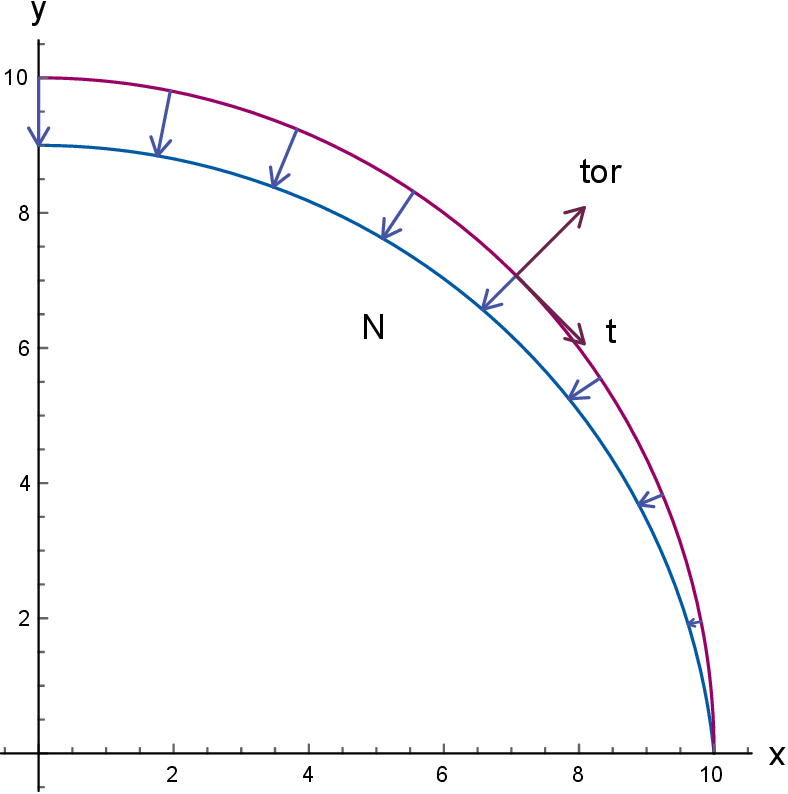}
\caption{Curved nanobeam with slider and roller supports: plot of the vector field $T\tort$.}
\label{fig6}
\end{figure}

\begin{figure}[htp]
\centering	
\includegraphics{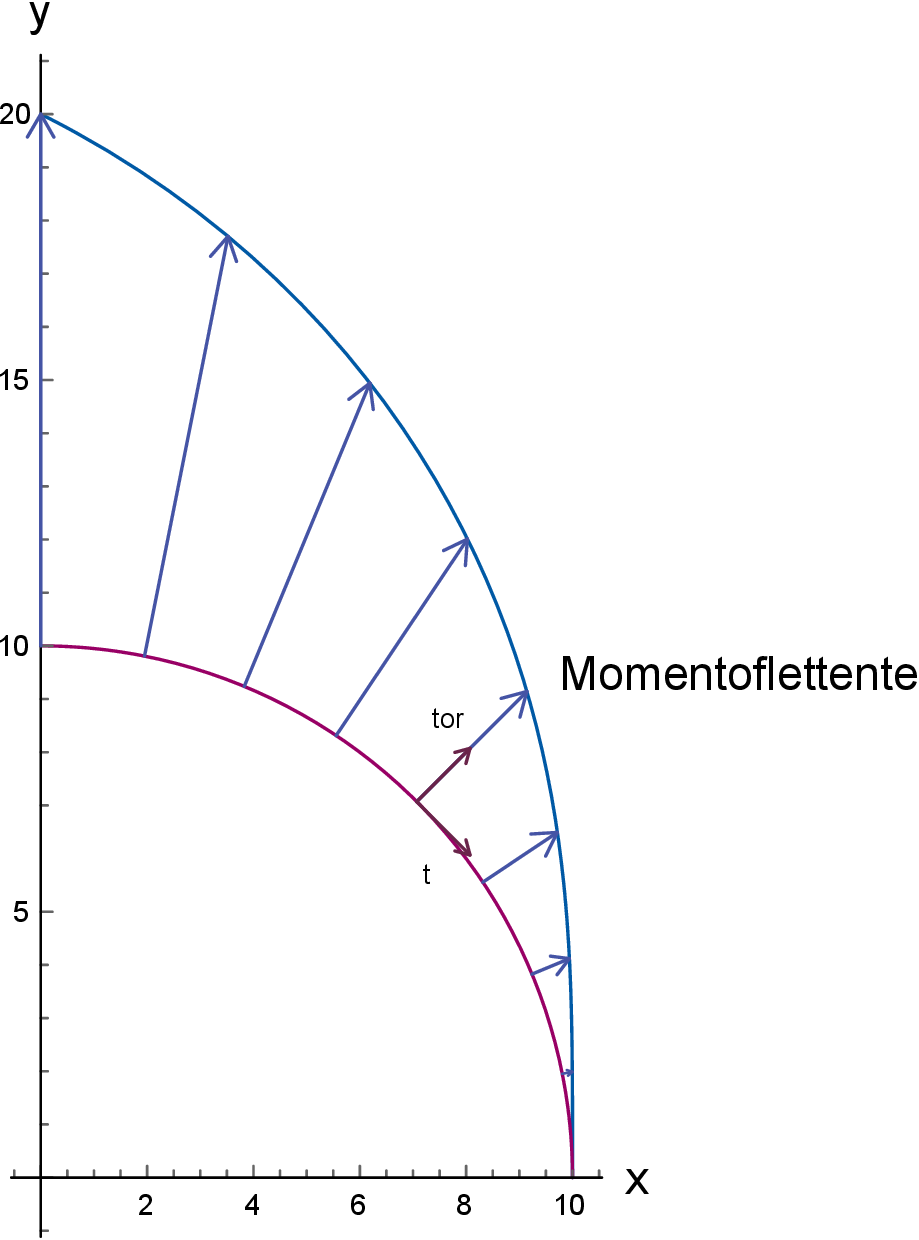}
\caption{Curved nanobeam with slider and roller supports: plot of the vector field $\R ($M$\k) = -M\tort$.}
\label{fig7}
\end{figure}

\begin{figure}
\centering
\psfragfig*[scale=0.52]{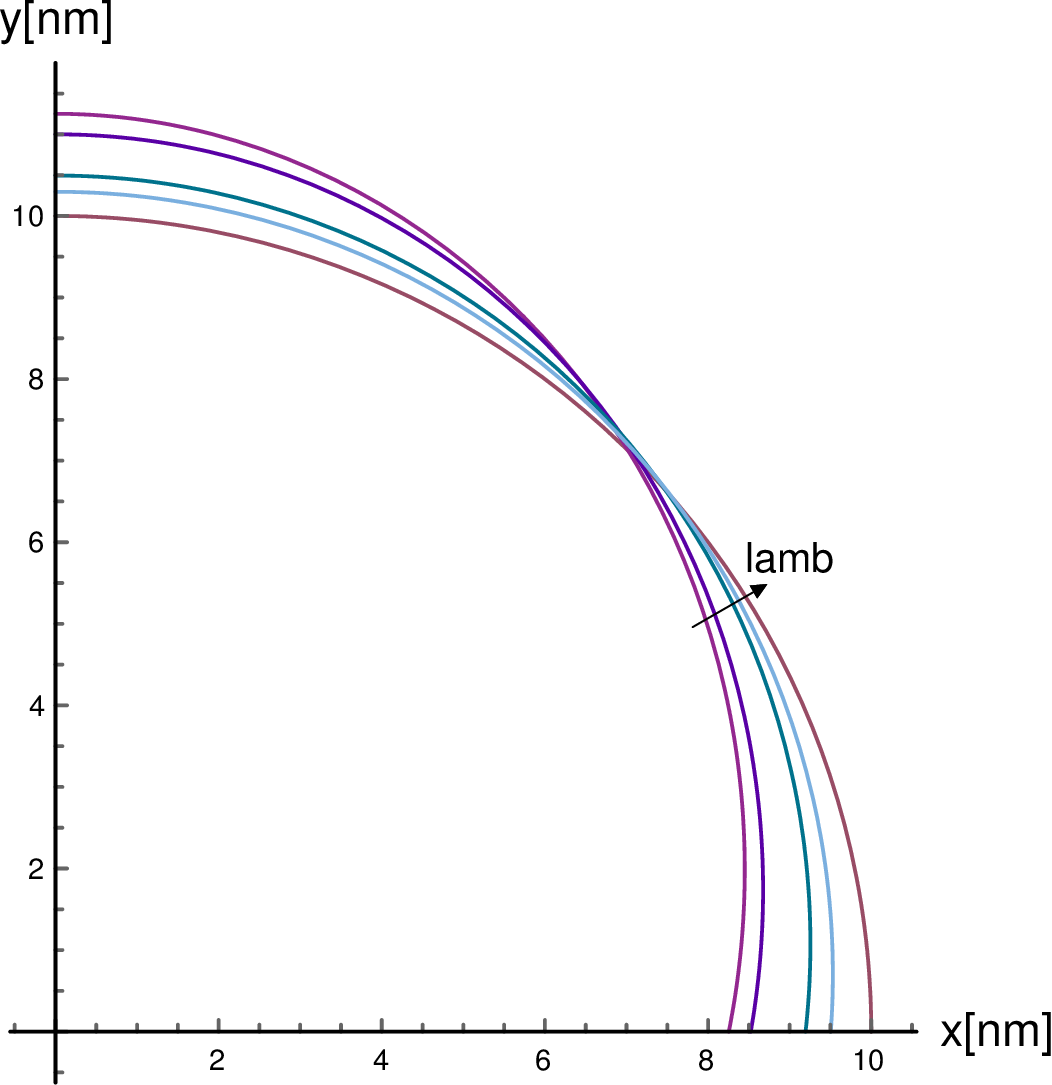}{
	\psfrag{x[nm]}{$x  [nm]$}
	\psfrag{y[nm]}{$y  [nm]$}
	\psfrag{lamb}{$\lambda$}
}
\caption{Curved nanobeam with slider and roller supports: deflection of the beam axis for $\,\lambda\in \{0; 0.1; 0.5; 1\}\,$.}
\label{def2}
\end{figure}

\begin{figure}[htp]
\centering	
\includegraphics{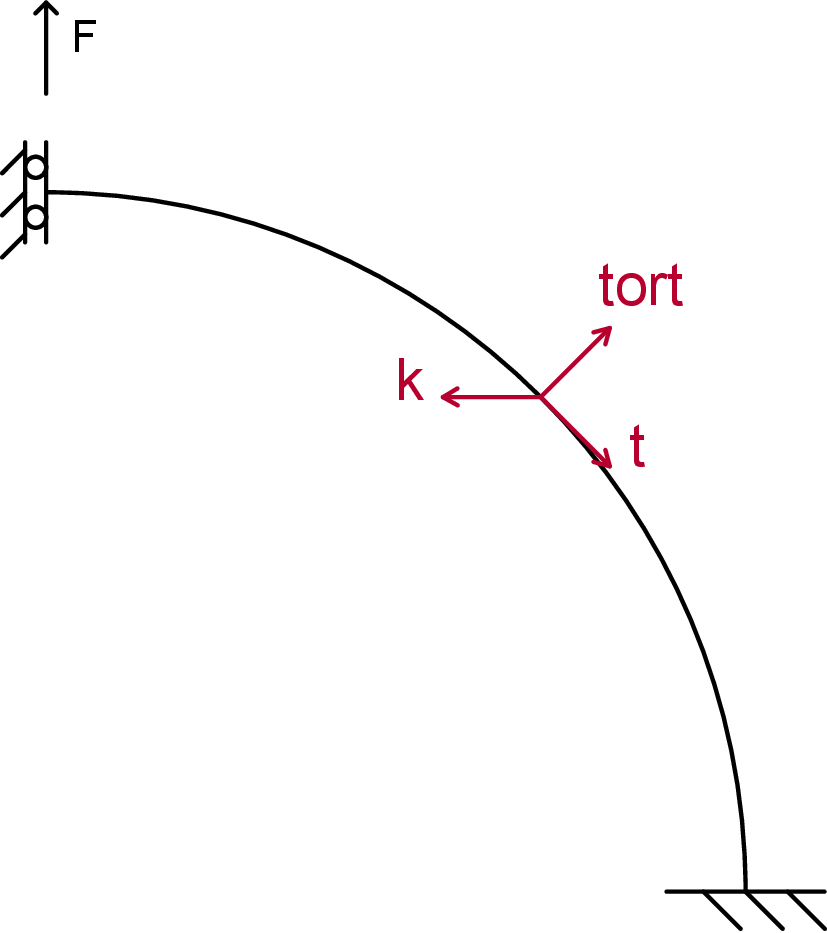}
\caption{Curved nanobeam with slider and clamp supports.}
\label{fig0b}
\end{figure}

\begin{figure}[htp]
\centering	
\includegraphics{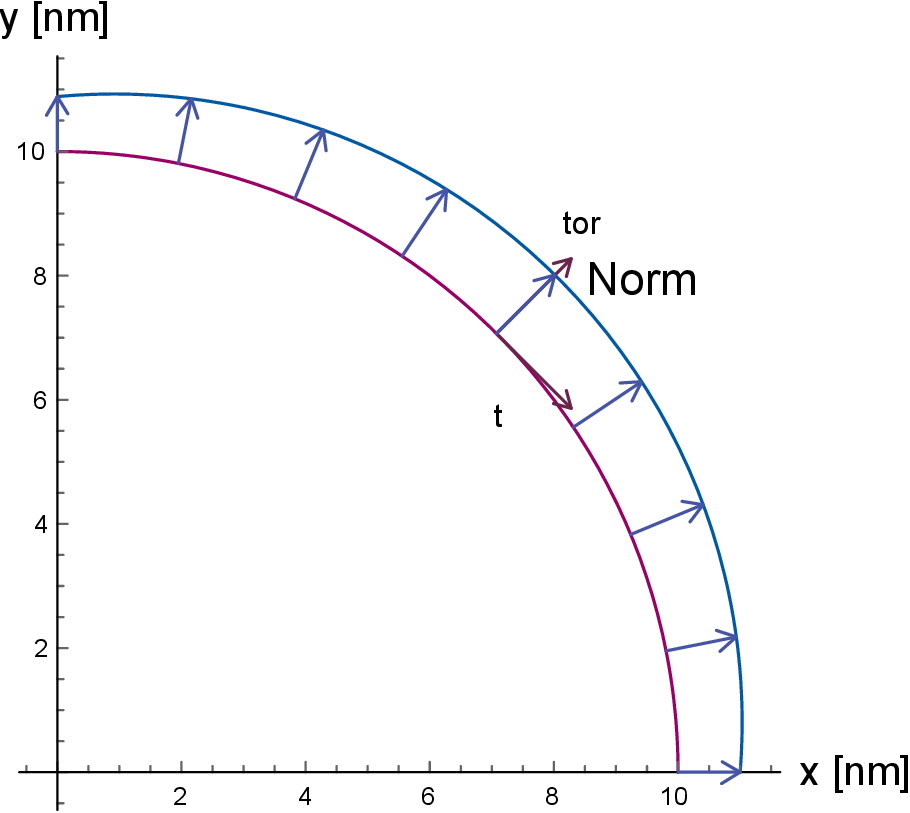}
\caption{Curved nanobeam with slider and clamp supports: plot of the vector field $N\tort$.}
\label{N3}
\end{figure}

\begin{figure}[htp]
\centering	
\includegraphics{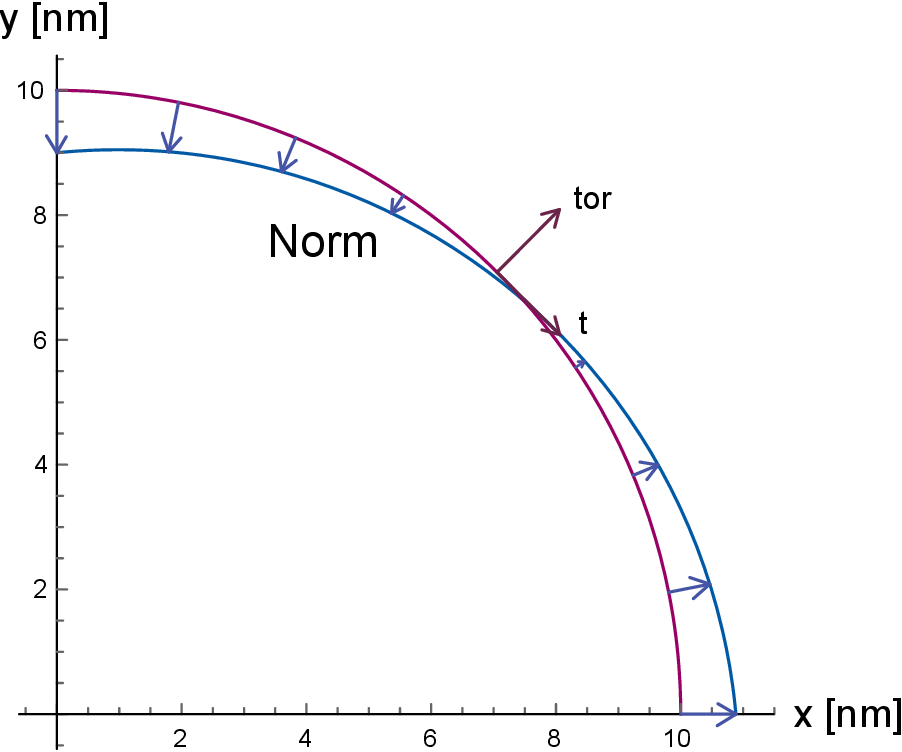}
\caption{Curved nanobeam with slider and clamp supports: plot of the vector field $T\tort$.}
\label{T3}
\end{figure}

\begin{figure}[htp]
\centering	
\includegraphics{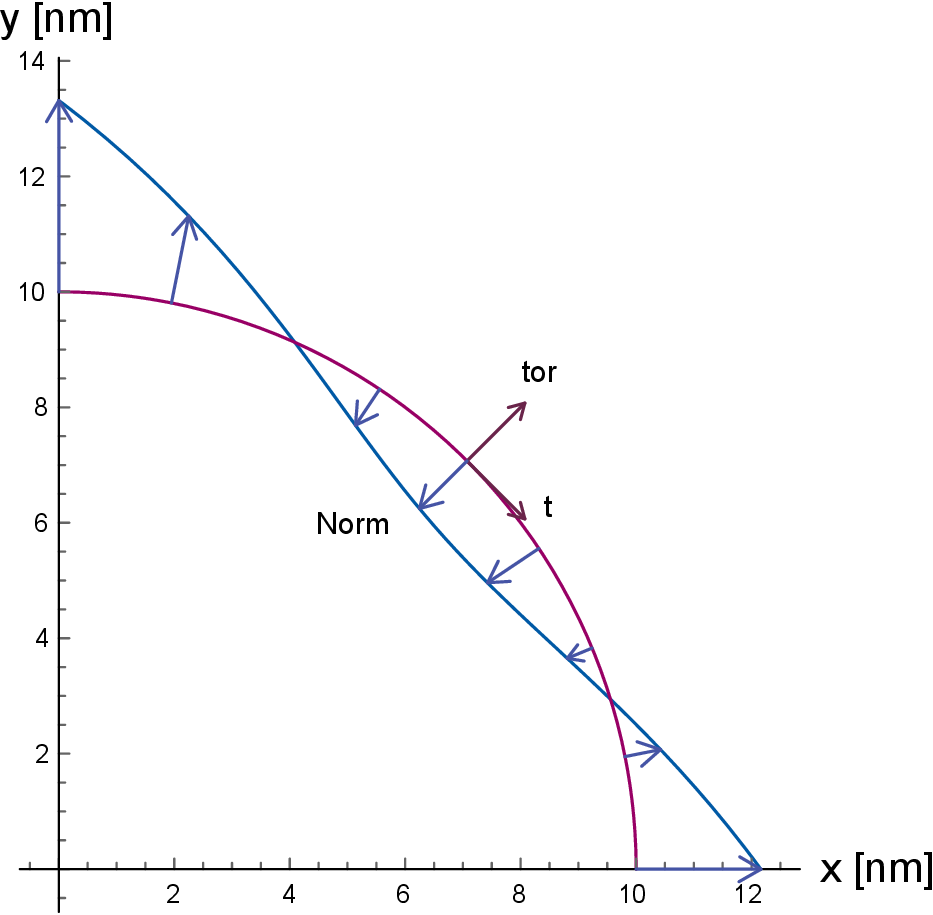}
\caption{Curved nanobeam with slider and clamp supports: plot of the vector field $\R ($M$\k) = -M\tort$.}
\label{M3}
\end{figure}

\begin{figure}[htp]
\centering	
\includegraphics{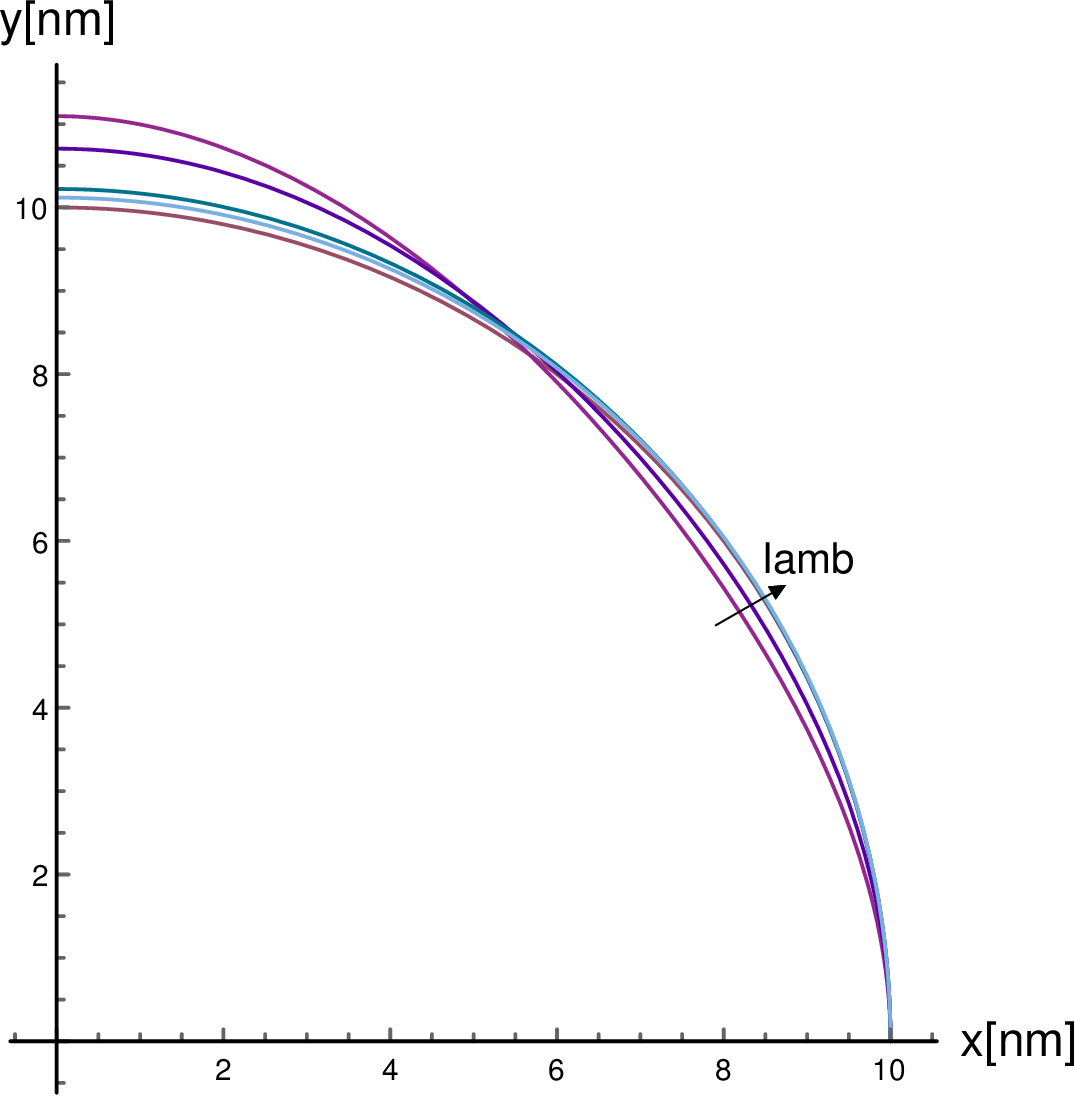}
\caption{Curved nanobeam with slider and clamp supports: deflection of the beam axis for $\lambda\in \{0; 0.1; 0.5; 1\}$.}
\label{defcaso3}
\end{figure}

\begin{figure}[htp]
\centering	
\includegraphics{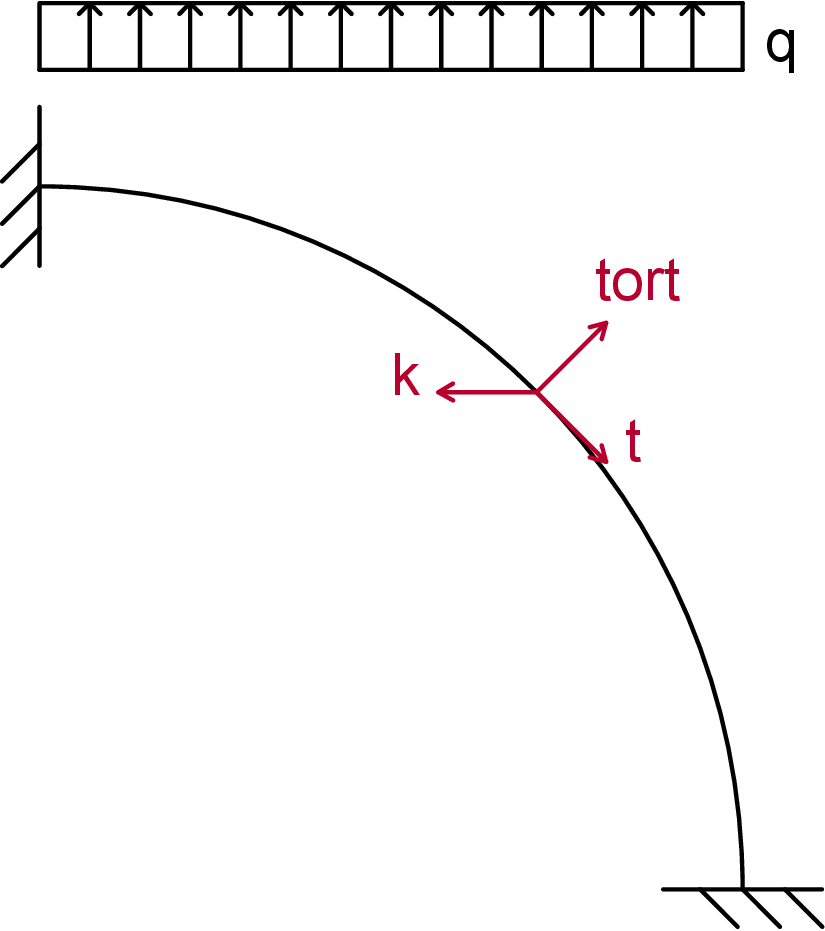}
\caption{Doubly clamped curved nanobeam}
\label{fig0c}
\end{figure}

\begin{figure}[htp]
\centering	
\includegraphics{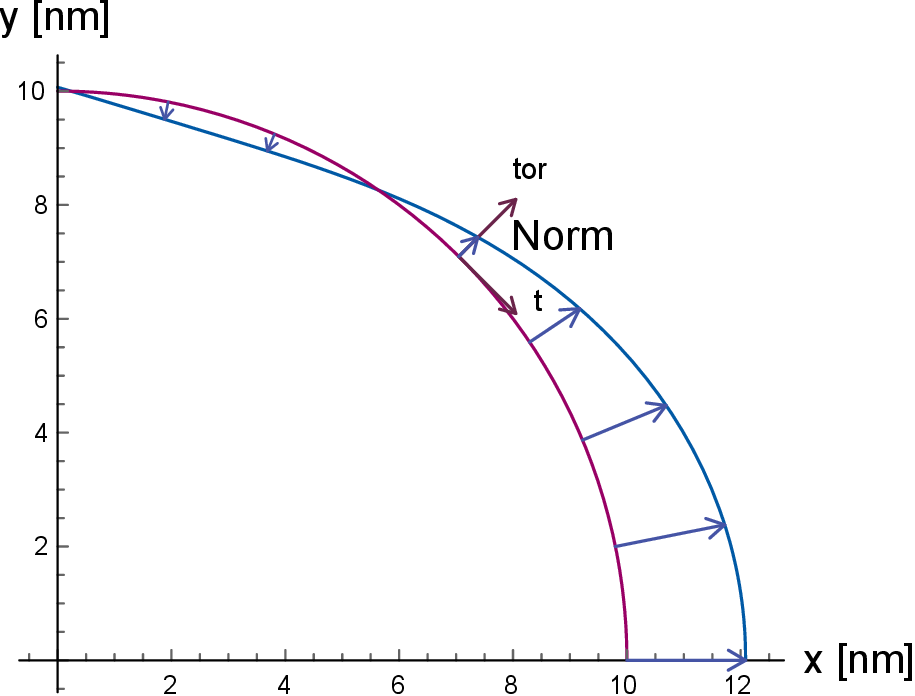}
\caption{Doubly clamped curved nanobeam: plot of the vector field $N\tort$.}
\label{N4}
\end{figure}

\begin{figure}[htp]
\centering	
\includegraphics{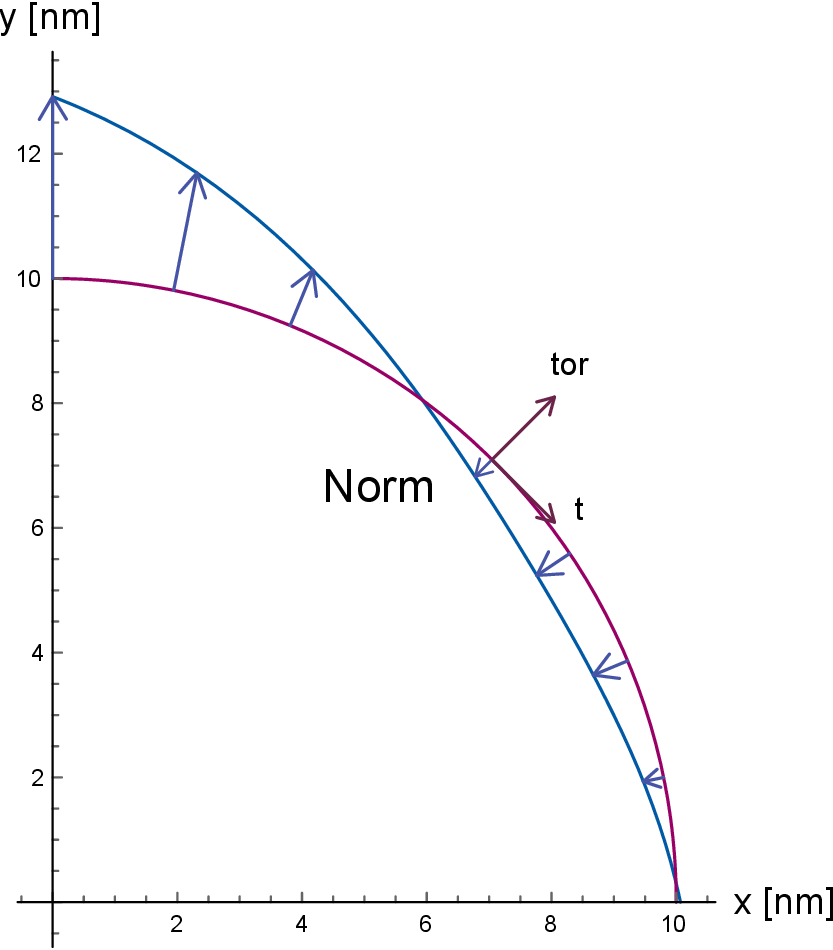}
\caption{Doubly clamped curved nanobeam: plot of the vector field $T\tort$.}
\label{T4}
\end{figure}

\begin{figure}[htp]
\centering	
\includegraphics{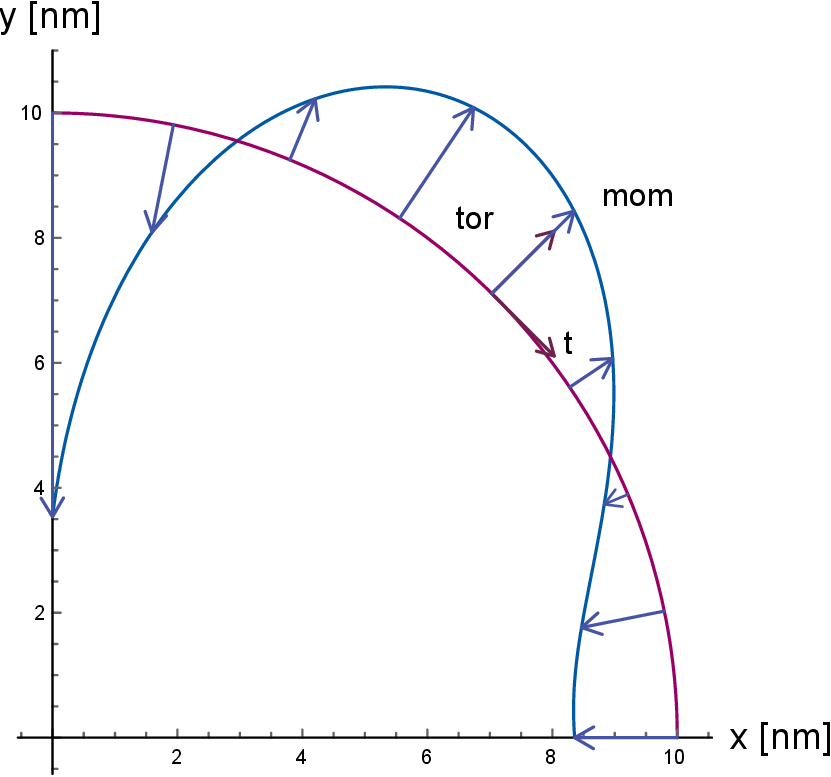}
\caption{Doubly clamped curved nanobeam: plot of the vector field $\,\R ($M$\k) = -M\tort$.}
\label{M4}
\end{figure}

\begin{figure}[htp]
\centering	
\includegraphics{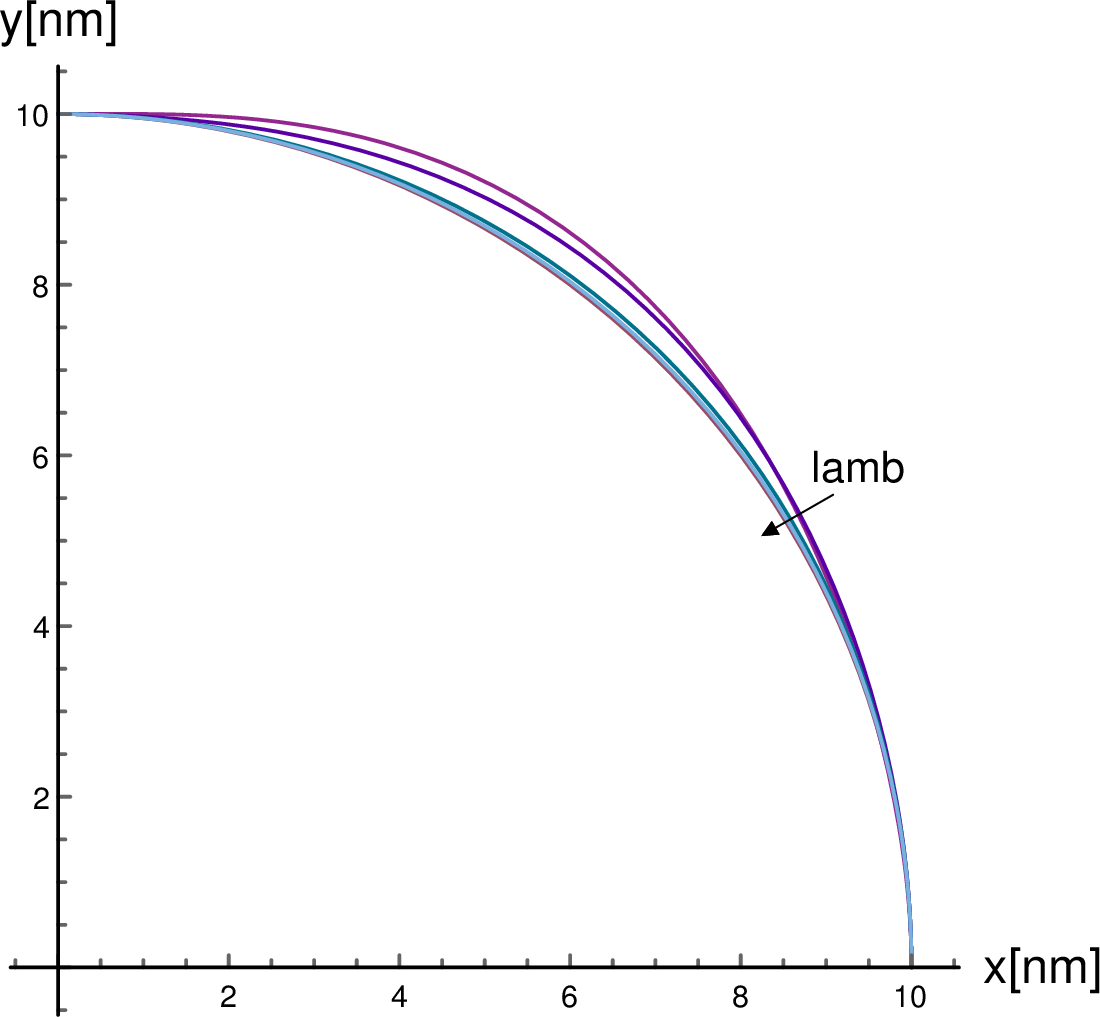}
\caption{Doubly clamped curved nanobeam: deflection of the beam axis for $\lambda\in\{0; 0.1; 0.5; 1\}$.}
\label{defcaso4}
\end{figure}

The solution procedure of the stress-driven nonlocal integral nanostructural problem, formulated in Sect.\ref{sec:NEP},
consists of three steps:
\begin{enumerate}
\item
\textbf{Step $\,1\,$.}
Solution of the differential static problem Eqs.\eqref{fm:DiffProbl} and \eqref{fm:StatBC}.
The interaction fields: axial force $\,N\,$, shear force $\,T\,$ and bending moment $\,M\,$
are expressed in terms of $\,n\,$ integration constants, with $\,n\,$ standing for redundancy degree.

For statically determinate beams we have that $\,n=0\,$ and the stress fields $\,N\,,T\,,M\,$ are
univocally determined by equilibrium requirements.
\item
\textbf{Step $\,2\,$.}
Evaluation of axial strain $\,\ee\,$ and bending curvature $\,\chi\,$ by making recourse to the
stress-driven integral convolutions Eq.\eqref{StressConv} or by solving the differential problem
Eq.\eqref{fm: Stressdifform} equipped with the nonlocal boundary conditions Eq.\eqref{fm: StressEqui},
based on the constitutive equivalency established in Prop.\ref{prop: nonlocal}.
\item
\textbf{Step $\,3\,$.}
Prescription of $\,3+n\,$ essential kinematic boundary conditions and
detection of the curved nanobeam nonlocal displacement $\,\Bu\,$ Eq.\eqref{fm:spost}.
\end{enumerate}
The size-dependent responses of the following four static schemes are studied.
\begin{enumerate}
\item
\textbf{Case $\,1\,$.}
Cantilever curved nanobeam subjected to a uniformly distributed vertical load $\,q = 0.1\,$ [nN]/[nm] 
directed upwards. 

The clamp is placed at the abscissa $\,\s = 0\,$, see Fig.\ref{fig0}.

The essential kinematic boundary conditions are
\begin{equation}
\Bu\di{0}=\Bo\in\VV\,, \qquad
\varphi\di{0}=0\in\Re\,.
\end{equation}
The natural static boundary conditions Eq.\eqref{fm:StatBC} take thus the form
\begin{equation}
\begin{cases}
N(L)&= 0\in\Re\,,\\
T(L)&= 0\in\Re\,,  \\
M(L)&= 0\in\Re\,.
\label{s.c.}
\end{cases}
\end{equation}

\break
The structural redundancy degree $\,n\,$ is vanishing and therefore the static problem, 
recalled at step $\,1\,$ above, provides the equilibrated
axial force $\,N\,$, shear force $\,T\,$ and bending moment $\,M\,$, whose graphic representations 
are reported in Figs.\ref{fig2}, \ref{fig3}, \ref{fig4}, respectively.
A schematic sketch of the bending axial vector $\,M\k\,$ and of the rotated vector $\,\R M\k\,$
is given in Fig.\ref{cross_sec}.
The nonlocal displacement field $\,\Bu\,$ is got by applying the steps $\,1\,$ and $\,2\,$ of the solution strategy above.
Plots of the deformed beam axis is provided in Fig.\ref{fig1} for increasing nonlocal parameter 
$\,\lambda\,$.
Both the deformed and undeformed beam axis are curved lines obtained by 
parameterizations in terms of the arch length $\,\s\,$ and plotted in a Cartesian plane $\,x\,$ [nm], $\,y\,$ [nm].
Note that the classical local solution of elasticity is obtained as the nonlocal parameter tends to zero 
($\lambda = 0^+$).
\item
\textbf{Case $\,2\,$.}
Curved nanobeam with slider and roller supports subjected to a force 
$\,F = 1\,$ [nN] directed upwards, concentrated at the abscissa $\,\s=0\,$, shown in Fig.\ref{fig0a}.
The essential kinematic boundary conditions are
\begin{equation}
\Bu\di{0}\punto\t\di{0}=0\in\Re\,, \qquad
\varphi\di{0}=0\in\Re\,, \qquad
\Bu\di{L}\punto\t\di{L}=0\in\Re\,.
\end{equation}
The natural static boundary conditions Eq.\eqref{fm:StatBC} take thus the form
\begin{equation}
\begin{cases}
T(0)&= -F\in\Re\,,\\
T(L)&= 0\in\Re\,,  \\
M(L)&= 0\in\Re\,.
\end{cases}
\end{equation}
The nanostructure is statically determinate, so that the redundancy degree $\,n\,$ is vanishing.
The solution method, applied to case $\,1\,$, provides scalar interactions  
$\,N\,$, $\,T\,$, $\,M\,$ and vector nonlocal displacements $\,\Bu\,$.
Graphic outputs are respectively presented in Figs.\ref{fig5}, \ref{fig6}, \ref{fig7} and \ref{def2}.
\item
\textbf{Case $\,3\,$.}
Curved nanobeam with slider and clamp supports subjected to a force 
$\,F = 10\,$ [nN] directed upwards, concentrated at $\,\s=0\,$, as shown in Fig.\ref{fig0b}.
The essential kinematic boundary conditions are
\begin{equation}
\Bu\di{0}\punto\t\di{0}=0\in\Re\,, \quad
\varphi\di{0}=0\in\Re\,, \quad
\Bu\di{L}=0\in\VV\,, \quad
\varphi\di{L}=0\in\Re\,.
\end{equation}
The natural static boundary conditions Eq.\eqref{fm:StatBC} take thus the form
\begin{equation}
T(0)= -F\in\Re\,.
\end{equation}
The nanostructure is statically indeterminate with redundancy degree $\,n=2\,$, so that
$\,3+2\,$ essential and $\,1\,$ natural boundary scalar conditions
have to be enforced.
The closure of the nonlocal structural problem is therefore guaranteed and the 
axial force $\,N\,$, shear force $\,T\,$, bending moment $\,M\,$ solution fields,  
associated with the scale parameter $\,\lambda=0.5\,$,
are depicted in Figs.\ref{N3}, \ref{T3}, \ref{M3}.
The nonlocal displacement solution field $\,\Bu\,$ is plotted in Fig.\ref{defcaso3} 
for increasing scale parameter $\,\lambda\,$.
\item
\textbf{Case $\,4\,$.}
Doubly clamped curved nanobeam under a uniformly distributed vertical load $\,q = 5\,$ [nN]/[nm] 
directed upwards, see Fig.\ref{fig0c}. 
The nanostructure is statically indeterminate with redundancy degree $\,n=3\,$,
so that the boundary conditions to be imposed are all of kinematic type
($\,2+1+2+1=6\,$ scalar prescriptions)
\begin{equation}
\Bu\di{0}=0\in\VV\,, \quad
\varphi\di{0}=0\in\Re\,, \quad
\Bu\di{L}=0\in\VV\,, \quad
\varphi\di{L}=0\in\Re\,.
\end{equation}
No natural static boundary conditions have to be inforced.
The elastostatic problem is solved to detect the static and kinematic nonlocal fields
$\,\set{N,T,M},\Bu\,$ in terms of the scale parameter $\,\lambda\,$.
The static outcomes $\,\set{N,T,M}\,$, corresponding to $\,\lambda=0.5\,$, 
are illustrated in Fig.\ref{N4}, \ref{T4}, \ref{M4},
while the nonlocal displacement vector field $\,\Bu\,$ is depicted in Fig.\ref{defcaso4}
for increasing scale parameter $\,\lambda\,$.
\end{enumerate}

\section{Results, discussion and closing remarks}
\label{sec:Concl}

The stress-driven nonlocal integral mechanics of straight elastic beams conceived in \citep{RB2017}
has been generalized in the present research to curved nanobeams, which are the principal
structural elements of new-generation Micro- and Nano-Electro-Mechanical Systems (M/NEMS).
The presented integral methodology has been shown to be equivalent to a convenient set of ordinary differential equations, equipped with non-standard nonlocal boundary conditions, generalizing thus to curved nanobeams
the constitutive equivalency proposition 7.1 in \citep{RB2017}.
The complex nonlocal elastic equilibrium equations governing the structural problems
of engineering interest have been therefore analytically simplified by transforming integral convolutions
into equivalent differential conditions.

\break
\noindent
Unlike strain-driven nonlocal integral and kinematically redundant couple stress models, 
the stress-driven nonlocal integral formulation of curved elastic nanostructures has been shown to lead to mathematically well-posed problems.
An effective solution procedure has been illustrated and implemented by Wolfram's software Mathematica
for a variety of nano-engineered structures exploited as sensors and actuators in M/NEMS applications.

In agreement with the experimental results collected in the paper by \citet{Abazari2015} exhibited by a wide range of advanced materials and small-scale devices, a stiffening nano-mechanical behaviour has been predicted by the stress-driven approach in all case-studies for increasing nonlocal parameter. 
The \textit{smaller-is-stiffer} phenomenon, also predicted in the recent research by
\citet{Fuschi2019}, seems to be a peculiar property in Nano-Engineering.

\bigskip\par\noindent\noindent
\textbf{Acknowledgements}\par

The financial support of the Italian Ministry for University and Research (P.R.I.N. National Grant 2017, 
Project code 2017J4EAYB; University of Naples Federico II Research Unit) is gratefully acknowledged.




\clearpage


\end{document}